\documentclass[a4paper,twocolumn]{article}

\usepackage{amsmath}
\DeclareMathOperator{\arctantwo}{arctan2}
\usepackage{hyperref}
\usepackage{graphicx}
\usepackage{color}

\newcommand{\be}{\begin{equation}}
\newcommand{\ee}{\end{equation}}
\newcommand{\bea}{\begin{eqnarray}}
\newcommand{\eea}{\end{eqnarray}}
\newcommand{\tr}{\mathrm{tr}}

\newcommand{\bfE}{\mathbf{E}}

\newcommand{\bfI}{\mathbf{I}}

\newcommand{\bfN}{\mathbf{N}}

\newcommand{\bfsigma}{\boldsymbol{\sigma}}
\newcommand{\bftau}{\boldsymbol{\tau}}

\newcommand{\bfzero}{\mbox{\boldmath $0$}}
\newcommand{\dd}{\;\!\mathrm{d}} 
\providecommand{\keywords}[1]
{
	\small	
	\textbf{\textit{Keywords---}} #1
}

\title{A novel class of electro-mechanical metamaterials for stress reduction through electric fields} 

\author{Mischa Blaszczyk\thanks{E-Mail: mischa.blaszczyk@rub.de. Ruhr-University Bochum, Institute of Mechanics of Materials, Universitätsstaße 150, 44801 Bochum, \underline{Germany}}        
	\and    
	Klaus Hackl\thanks{E-Mail: klaus.hackl@rub.de. Ruhr-University Bochum, Institute of Mechanics of Materials, Universitätsstaße 150, 44801 Bochum, \underline{Germany}}}%

\date{\today}

\begin{document}
\maketitle

\begin{abstract}
While most previous developed metamaterials only consider a single physical effect, we introduce a novel class of electro-mechanical metamaterials, which allows a direct controllable reduction of the total stress by applying an electric field counteracting the mechanical stress. The solution of the resulting minimization problem yields a relation involving the eigenvalues of the mechanical stress tensor. Additionally, we evaluate the constrained cases allowing only tensile or compressive stresses, respectively, and consider the plane stress problem. We show numerical results for all cases and discuss, to what extent a stress reduction is possible.
\end{abstract}

\keywords{metamaterials, electro-mechanical coupling, stress reduction, Maxwell stress tensor}

\section{Introduction}\label{sec:Introduction}
Metamaterials are artificially created multiscale materials, which may possess properties not found in nature. In contrast to specific length scales, differentiation bet\-ween the scales is done by considering the microscale as consisting of an array of (periodic) unit cells, whose precise geometry and arrangement enable the often unusual physical effects observable in metamaterials, while the macroscale refers to the scale where these effects emerge. The material properties of the metamaterial can significantly differ from the original components it is made from. An overview and the state of the art can be found e.g. in the review articles \cite{smith2004metamaterials,liu2011metamaterials}.

Examples of electromagnetic metamaterials include substances with simultaneously negative permittivity and permeability \cite{veselago}, metamaterial absorbers \cite{landy2008perfect}, artificial magnetism \cite{ginn2012realizing}, negative refraction index \cite{valentine2008three}, topological insulators \cite{khanikaev2013photonic}, electromagnetic waveguides \cite{baena2005near} and performance augmented antennas \cite{dong2012metamaterial}. Metamaterials are not limited to electromagnetism. Acoustical, optical and mechanical metamaterials are also of great research interest and oftentimes show analogies to electromagnetic metamaterials. Examples for proposed or observed phenomenons are acoustic wave\-guides \cite{garcia2012quasi}, bandgaps \cite{huang2010band}, optical super lenses \cite{fang2002imaging}, seismic metamaterials \cite{brule2014experiments, colombi2016seismic}, materials with negative bulk modulus and negative mass density \cite{dingliu}, auxetic materials, i.e. materials with negative Poisson's ratio \cite{ren2018auxetic}, ultra lightweight but very strong materials \cite{meza2015resilient} and metamaterial cloaking \cite{greenleaf2012cloaked}.

\begin{figure}[h!]
	\centering
	\includegraphics[width=0.45\textwidth]{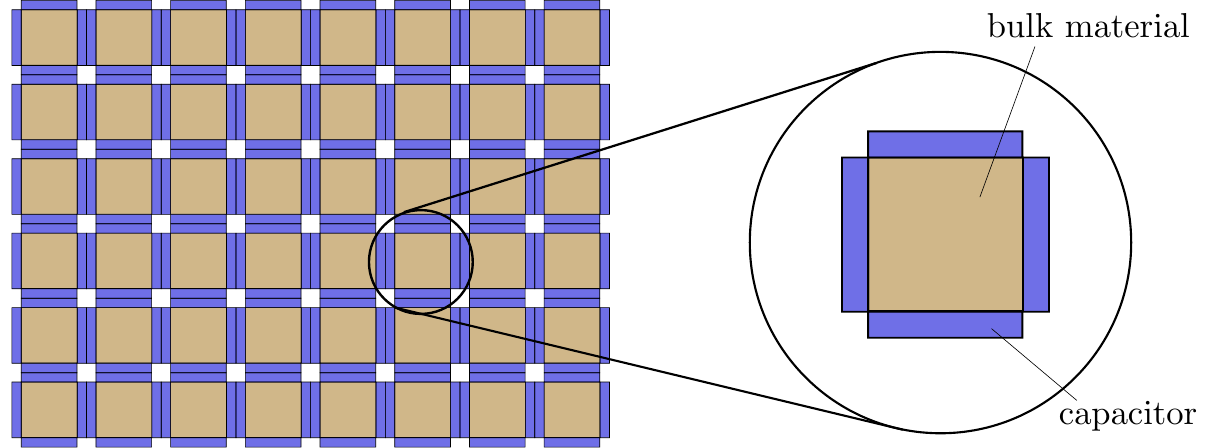}
	\caption{\label{fig0}Conceptional periodic microscale arrangement (left) and unit cell (right) in two dimensions.}
\end{figure}

While most metamaterials only consider a single physical effect (e.g. acoustic or electric metamaterials), in this paper we propose a novel class of theoretical metamaterials considering mechanical and electric effects. The key idea is to combine insulating and conducting materials, with the aim to directly control the total stress of the insulating material by applying an electric field, generated by the conducting material, counteracting the mechanical stress. A unit cell of our material could be constructed e.g. by surrounding an insulating material cube by a (conducting) capacitor in each spatial direction. Figure \ref{fig0} shows a possible periodic array in two dimensions. The electric field between two capacitor plates may be assumed constant and the combination of capacitors in each spatial direction ensures that the direction of the resulting electric field can be controlled arbitrarily. For sufficiently small unit cell sizes, the mechanical stress within the bulk material of a single unit cell may also assumed to be constant and a stress reduction would be possible. 

Some materials show a very different behavior for tensile and compressive loads, e.g. concrete \cite{doi:10.1680/macr.1968.20.65.221}, which is important in many engineering applications. Therefore, we also investigate the minimization problem with the additional constraint, that only tensile or compressive stresses, respectively, are allowed. Additionally, we discuss the two-dimensional case of plane stress. We obtain mappings related to the eigenvalues for the different problems. Lastly, we discuss our findings and to what extent the mechanical stress can be reduced.
\section{Material model}\label{sec:Material}

\subsection{Stress minimization}\label{sec:Material1}
We start by assuming the stress in our insulating bulk material depends on the mechanical stress tensor
\be
\bfsigma = \begin{bmatrix} \sigma_{xx} & \sigma_{xy} & \sigma_{xz} \\ \sigma_{xy} & \sigma_{yy} & \sigma_{yz} \\ \sigma_{xz} & \sigma_{yz} & \sigma_{zz}\end{bmatrix}
\ee
and the Maxwell stress tensor for linear materials \cite{dorfmann2014nonlinear}
\be
\bftau^m = \varepsilon_0 (\varepsilon_r \bfE \otimes \bfE - \frac{1}{2} (\bfE \cdot \bfE) \; \bfI) \text{,}
\ee
both written in tensor notation. Here, $\varepsilon_0$ is the vacuum permittivity, $\varepsilon_r$ is the relative permittivity of the material and $\bfE$ is the electric field. The symbol $\bfI$ denotes the identity tensor. Thus, the total stress $\bfsigma^t$ of our material can be written as
\be
\bfsigma^t = \bfsigma + \bftau^m \text{,}
\ee
resulting in a symmetric tensor, as both $\bfsigma$ and $\bftau^m$ are symmetric. In three dimensions, the mechanical stress tensor consists of six independent components, while the electric field only possesses three independent components. Therefore, it is immediately clear, that a perfect stress absorption (i.e. $\bfsigma^t = \bfzero$) is only possible for rare special cases. However, the best possible reduction - choosing $\bfE$ such that $||\bfsigma^t||^2$ becomes minimal - can still be calculated. Here, $||\cdot||$ is the Frobenius norm. The minimization problem reads
\be
||\bfsigma^t||^2 = ||\bfsigma + \bftau^m||^2 \rightarrow \underset{\bfE}{\min} \text{.} \label{eqmin}
\ee
The necessary condition for the minimization is
\be
\nabla_{\bfE} || \bfsigma + \bftau^m||^2  \overset{!}{=} \bfzero \text{,} \label{eqmincond}
\ee
where $\nabla$ is the Nabla operator. To solve the problem, we write down at first the Maxwell stress tensor in index notation using the Einstein summation convention as
\be
\tau_{ij}^m = \varepsilon_0 (\varepsilon_r E_i E_j - \frac{1}{2} E_k E_k \delta_{ij} ) \text{.}
\ee
Then, we calculate its derivative with respect to the electric field
\be
\frac{\partial \tau_{ij}^m}{\partial E_e} = \varepsilon_0 (\varepsilon_r \delta_{ie} E_j + \varepsilon_r E_i \delta_{je} - \frac{1}{2} 2 \delta_{ke} E_k \delta_{ij}) \text{.} \label{eqderiv}
\ee
Using symmetry of the included tensors, Eq. \eqref{eqmincond} yields in index notation
\be
\frac{\partial (||\sigma_{ij} + \tau^m_{ij}||^2)}{\partial E_e} = 2 (\sigma_{ij} + \tau^m_{ij}) \frac{\partial \tau^m_{ij}}{\partial E_e} = 0 \text{.} 
\ee
Inserting the derivative (Eq. \eqref{eqderiv}) and dividing by $2 \varepsilon_0$, we obtain
\bea
\varepsilon_0 (\varepsilon_r E_i E_j - \frac{1}{2} E_k E_k \delta_{ij})(\varepsilon_r \delta_{ie}E_j+\varepsilon_r \delta_{je} E_i \nonumber \\ - \delta_{ij} E_e) + \sigma_{ij} (\varepsilon_r \delta_{ie}E_j + \varepsilon_r \delta_{je} E_i - \delta_{ij} E_e) = 0 \text{,} 
\eea
which can be simplified to
\bea
\varepsilon_0 (2 \varepsilon_r^2-\varepsilon_r) E_e E_j E_j + \varepsilon_0 (\frac{3}{2} - \varepsilon_r) (E_k E_k E_e) \nonumber \\
+ \varepsilon_r \sigma_{ej} E_j + \varepsilon_r \sigma_{ie} E_i - \sigma_{ii} E_e = 0 \text{.}
\eea
In this form, the result can be transformed back into tensor notation:
\be
\varepsilon_0 |\bfE|^2 \bfE (2 \varepsilon_r^2 - 2 \varepsilon_r + \frac{3}{2}) + 2 \varepsilon_r \bfsigma \cdot \bfE - \tr \bfsigma \bfE = \bfzero \text{.} \label{eqtens}
\ee
In Eq. \eqref{eqtens}, $|\cdot|$ denotes the Euclidean norm. We rearrange the equation to obtain the eigenvalue problem for the mechanical stress tensor $\bfsigma$:
\be
\bfsigma \cdot \bfE = (\frac{1}{2 \varepsilon_r} \tr \bfsigma - (\varepsilon_r - 1 + \frac{3}{4 \varepsilon_r}) \varepsilon_0 |\bfE|^2) \; \bfE \label{eqsmr} \text{.}
\ee 
Thus it follows that in order to minimize the total stress, the electric field needs to have the form $\bfE = \alpha \bfN$, where $\bfN$ is a unit eigenvector of $\bfsigma$ and $\alpha := |\bfE|$ can be calculated from Eq. \eqref{eqsmr}, denoting the stress tensor eigenvalues $\lambda_i$, as
\be
\lambda_i = \frac{1}{2 \varepsilon_r} \tr \bfsigma - (\varepsilon_r - 1 + \frac{3}{4 \varepsilon_r}) \varepsilon_0 |\bfE|^2 \text{,}
\ee
resulting in
\be
\alpha = |\bfE| = \sqrt{\frac{\tr \bfsigma - 2 \varepsilon_r \lambda_i}{2 \varepsilon_0 (\varepsilon_r^2-\varepsilon_r+\frac{3}{4})}} \text{.} \label{eqalpha}
\ee
It should be noted here, that - depending on the specific stress state and the parameter $\varepsilon_r$ - a real solution cannot always be found. In order to obtain a minimized solution for the problem, $|\bfE|$ would be required to become imaginary, which is unphysical.

In the remainder of this paper, we will restrict ourselves to the case of non-polarizable materials, i.e. $\varepsilon_r = 1$. While the influence of $\varepsilon_r$ may be significant for the results, a parameter study would go beyond the scope of this contribution. It should be noted, that the eigenvalues of the Maxwell stress tensor then always take the form \cite{hoffmann}
\be
\lambda_\tau = \{-\lambda_{m}, -\lambda_{m}, +\lambda_{m}\} \text{,} \label{eqevmws0}
\ee
with
\be
\lambda_{m} = \varepsilon_0 \frac{|\bfE|^2}{2}\ge 0 \text{.} \label{eqevmws}
\ee
In the following, we use the convention $\lambda_3 \ge \lambda_2 \ge \lambda_1$. Using the results of the minimization, the absolute value of the Maxwell stress tensor eigenvalues depends on the eigenvalues of the mechanical stress as
\bea
\lambda_{m,1} &=& \frac{1}{3} (-\lambda_1+\lambda_2+\lambda_3) \text{,} \nonumber \\ \lambda_{m,2} &=& \frac{1}{3} (\lambda_1-\lambda_2+\lambda_3) \text{,} \nonumber \\ \lambda_{m,3} &=& \frac{1}{3} (\lambda_1+\lambda_2-\lambda_3) \text{,}
\eea
which can be proven by calculating the corresponding value for $\alpha$ from Eq. \eqref{eqalpha} and inserting the result into Eq. \eqref{eqevmws}, using the principal axes form of $\bfsigma$. The index number corresponds to the specific eigenvalue of the stress tensor. With respect to the original coordinate system, the Maxwell stress tensor then takes the form
\be
\bftau^m = \varepsilon_0 (\alpha^2 \bfN \otimes \bfN - \frac{1}{2} \alpha^2 \bfI) \text{.}
\ee
Again considering the principal axes, we now calculate the relative stress reduction 
\be
\sigma_{\mathrm{rel}} := \frac{||\bfsigma + \bftau^m||}{||\bfsigma||} \text{,} \label{eqsigmarel}
\ee
depending on the eigenvalues of the mechanical stress tensor. Using the expression
\bea
||\bfsigma + \bftau^m||^2 = ||\bfsigma||^2 + 2 \tr (\bfsigma \cdot \bftau^m) + ||\bftau^m||^2 \nonumber \\
= (\lambda_1^2 + \lambda_2^2 + \lambda_3^2) + 2 (\lambda_1\lambda_{m} - \lambda_2\lambda_{m} - \lambda_3\lambda_{m}) \nonumber \\
+ \frac{1}{3} (-\lambda_1+\lambda_2+\lambda_3)^2 \text{,}
\eea
we obtain 
\bea
\sigma_{\mathrm{rel},1} &= \sqrt{\frac{2}{3} (1+\displaystyle\frac{\lambda_1\lambda_2+\lambda_1\lambda_3-\lambda_2\lambda_3}{\lambda_1^2+\lambda_2^2+\lambda_3^2})} \text{,}  \\
\sigma_{\mathrm{rel},2} &= \sqrt{\frac{2}{3} (1+\displaystyle\frac{\lambda_1\lambda_2-\lambda_1\lambda_3+\lambda_2\lambda_3}{\lambda_1^2+\lambda_2^2+\lambda_3^2})} \text{,}  \\
\sigma_{\mathrm{rel},3} &= \sqrt{\frac{2}{3} (1+\displaystyle\frac{-\lambda_1\lambda_2+\lambda_1\lambda_3+\lambda_2\lambda_3}{\lambda_1^2+\lambda_2^2+\lambda_3^2})} \text{.} 
\eea
From this result we are able to proof that choosing the smallest eigenvalue of $\bfsigma$, $\lambda_1$, yields the optimal solution, i.e. $\sigma_{\mathrm{rel}}$ becomes minimal compared to the other eigenvalues. A detailed derivation can be found in the supplemental material \cite{suppl}, which also includes additional details regarding the constrained minimization and some simple numerical examples. Finally, we note that since our starting point (Eq. \eqref{eqmin}) uses the Frobenius norm, which only attains values in the range $\ge 0$ and will approach infinity for $|\bfE| \rightarrow \infty$, the calculated critical point $\lambda_{m,1}$ is the global minimum of our function.
\subsection{Tensile stress}\label{sec:Material2}
In the remainder of this paper, we only consider the principal axes system, since as we have shown previously we are always able to transform our problem to a coordinate system with respect to the principal axes. Our aim here is to evaluate, whether a stress minimization is possible when only tensile stresses are allowed to occur and how the eigenvalues of the Maxwell stress tensors have to be calculated to obtain the minimal possible stress. Then, inserting this result into Eq. \eqref{eqevmws} yields the electric field necessary for this task. 

Due to the structure of the Maxwell stress tensor and its eigenvalues Eq. \eqref{eqevmws0}, it is immediately clear that this task is impossible, if $\bfsigma$ possesses two or three negative eigenvalues. We will discuss the other cases in detail. For the case of tensile stresses, we show the complete calculations here once. A detailed derivation for all cases, following the same procedure, can be found in the supplemental material \cite{suppl}. First, we assume 
\be
\bfsigma = \begin{bmatrix}\lambda_1 & 0 & 0 \\ 0 & \lambda_2 & 0 \\ 0 & 0 & -\lambda_3\end{bmatrix} \text{,}
\ee
with $\lambda_1 \ge \lambda_2 \ge \lambda_3 \ge 0$. We obtain the minimization problem
\be
||\bfsigma + \bftau^m||^2 \rightarrow \underset{\lambda_{m}}{\min} \text{,}
\ee
with the inequality conditions
\bea
\lambda_2 \ge \lambda_{m} \quad \text{and} \quad \lambda_{m} \ge \lambda_{3} \text{.}
\eea
Incorporating these into our minimization problem, we obtain the Lagrangian
\bea
L(\lambda_{m},\alpha,\beta) = (\lambda_1-\lambda_{m})^2 + (\lambda_2 - \lambda_{m})^2 \nonumber \\ + (-\lambda_3 + \lambda_{m})^2 - \alpha g(s) - \beta h(t) \rightarrow \underset{\lambda_{m}, \alpha, \beta}{\min} \; \text{,}
\eea
Here, $s$ and $t$ are slack variables and $\alpha$ and $\beta$ are Lagrange multipliers. The functions $g(s)$ and $h(t)$ contain the constraint information of the problem:
\bea
g(s) &=& \lambda_{m} - \lambda_2 + s^2 \; \text{,}\nonumber \\
h(t) &=& \lambda_3 - \lambda_{m} + t^2 \; \text{,}
\eea
We obtain the derivatives
\bea
\frac{\partial L}{\partial \lambda_{m}} =& -2(\lambda_1 - \lambda_{m}) - 2(\lambda_2 - \lambda_{m}) + 2 (\lambda_{m} - \lambda_3) \nonumber \\
& - \alpha + \beta = 0 \text{,} \label{eqlagr} \\
-\frac{\partial L}{\partial \alpha} =& g(s) = \lambda_{m} - \lambda_2 + s^2 = 0 \text{,} \label{eqlagra} \\
-\frac{\partial L}{\partial \beta} =& h(t) = \lambda_3 - \lambda_{m} + t^2 = 0 \text{,} \label{eqlagrb}
\eea
and Karush-Kuhn-Tucker (KKT) conditions
\be
\alpha s = 0 \quad \text{and} \quad \beta t = 0 \text{,}
\ee
from which the four cases
\bea
\text{(1)} \quad & \alpha = \beta = 0, & s^2 > 0, t^2 > 0 \text{,} \nonumber \\
\text{(2)} \quad & \alpha \ne 0, \beta = 0, & s^2 = 0, t^2 > 0 \text{,} \nonumber \\
\text{(3)} \quad & \alpha = 0, \beta \ne 0, & s^2 > 0, t^2 = 0 \text{,} \nonumber \\
\text{(4)} \quad & \alpha \ne 0, \beta \ne 0, & s^2=t^2 = 0 \text{.}
\eea
follow. We use this notation for all following examples as well. We start with case (1). As both Lagrange multipliers vanish, from Eq. \eqref{eqlagr} we obtain the solution
\be
\lambda_{m} = \frac{1}{3}(\lambda_1 + \lambda_2 + \lambda_3) \text{.} \label{eqt1}
\ee
Incorporating Eqs. \eqref{eqlagra} and \eqref{eqlagrb}, our result is only valid if both of the following inequalities hold:
\be
2\lambda_2 - \lambda_1 - \lambda_3 \ge 0 \quad \text{and} \quad \lambda_1+\lambda_2-2\lambda_3 \ge 0 \text{.} \label{eqt2}
\ee
For case (2), from Eq. \eqref{eqlagra} we calculate
\be
\lambda_{m} = \lambda_2 \text{.} \label{eqt3}
\ee
Eq. \eqref{eqlagrb} requires
\be
t^2 = \lambda_2 - \lambda_3 \ge 0 \text{,}
\ee
which is automatically fulfilled, as we demanded $\lambda_2 \ge \lambda_3$. Similarly, for case (3) from Eq. \eqref{eqlagrb} we obtain
\be
\lambda_{m} = \lambda_3 \text{.} 
\ee
This solution will always be equal or worse than the previous one. Finally, for case (4), we obtain the limit case $\lambda_{m} = \lambda_{2} = \lambda_{3}$. Thus in conclusion, if possible (i.e. both inequalities are fulfilled), the solution $\lambda_{m} = \frac{1}{3}(\lambda_1 + \lambda_2 + \lambda_3)$ should always be chosen, elsewise $\lambda_{m} = \lambda_2$. This can be seen by inserting both solutions in the Lagrangian and calculating the difference between the first and second solution, which returns the expression $-\frac{1}{2}(\lambda_1-2\lambda_2+\lambda_3)^2 \le 0$, implying a global minimum for the first solution. Analogously, this can also be done for the following examples.

Next, we consider the case
\be
\bfsigma = \begin{bmatrix}\lambda_1 & 0 & 0 \\ 0 & \lambda_2 & 0 \\ 0 & 0 & \lambda_3\end{bmatrix} \text{,}
\ee
with $\lambda_1 \ge \lambda_2 \ge \lambda_3 > 0$. Thus, the maximum allowed solution is $\lambda_m = \lambda_2$, otherwise we would obtain at least one negative eigenvalue for the total stress (cf. Eq. \eqref{eqevmws0}). The Lagrangian is then 
\bea
L(\lambda_m,\alpha) &=& (\lambda_1-\lambda_m)^2+(\lambda_2-\lambda_m)^2+(\lambda_3+\lambda_m)^2 \nonumber \\
& &- \alpha g(s) \rightarrow \underset{\lambda_{m}, \alpha}{\min} \text{,}
\eea
with derivatives
\bea
\frac{\partial L}{\partial \lambda_{m}} =& -2(\lambda_1 - \lambda_{m}) - 2(\lambda_2 - \lambda_{m}) + 2 (\lambda_{3} + \lambda_m) \nonumber \\
&- \alpha = 0 \text{,} \label{eqlagr2} \\
-\frac{\partial L}{\partial \alpha} =& g(s) = \lambda_{m} - \lambda_2 + s^2 = 0 \text{,} \label{eqlagra2}
\eea
and the KKT-condition
\be
\alpha s = 0 \text{,}
\ee
from which follow the two cases
\bea
\text{(1)} \quad & \alpha = 0, & s^2 > 0 \text{,} \nonumber \\
\text{(2)} \quad & \alpha \ne 0, & s^2 = 0 \text{.}
\eea
From case (1) we calculate
\be
\lambda_m = \frac{1}{3} (\lambda_1 + \lambda_2 - \lambda_3) \text{,}
\ee
using Eq. \eqref{eqlagr2}, if $\lambda_m \le \lambda_2$ holds (Eq. \eqref{eqlagra2}). Elsewise, case (2) yields
\be
\lambda_m = \lambda_2 \text{.}
\ee
\subsection{Compressive stress}
Analogously to the previous tensile minimization, we now consider the minimization problem with the constraint, that only compressive stresses remain in the material, by evaluating our equations in the coordinate system with respect to the principal axes. If $\bfsigma$ only contains positive eigenvalues, no solution can be found. To start, we assume
\be
\bfsigma = \begin{bmatrix}\lambda_1 & 0 & 0 \\ 0 & \lambda_2 & 0 \\ 0 & 0 & -\lambda_3\end{bmatrix}\text{,}
\ee
with $\lambda_1, \lambda_2, \lambda_3 > 0$ and $\lambda_3 \ge \lambda_1$. The maximum allowed solution is $\lambda_m = \lambda_3$. Additionally, we require $\lambda_m \ge \lambda_1$. The Lagrangian is
\bea
L(\lambda_m,\alpha,\beta) = (\lambda_1-\lambda_m)^2+(\lambda_2-\lambda_m)^2+ \nonumber \\ (\lambda_m-\lambda_3)^2 - \alpha g(s) - \beta h(t) \rightarrow \underset{\lambda_{m}, \alpha, \beta}{\min} \text{.}
\eea
We calculate
\be
\lambda_{m} = \frac{1}{3} (\lambda_1 + \lambda_2 + \lambda_3) \text{,}
\ee
if both of the following inequalities hold
\be
\lambda_2 + \lambda_3 - 2 \lambda_1 \ge 0 \quad \text{and} \quad 2 \lambda_3 - \lambda_1 - \lambda_2 \ge 0 \; \text{.}
\ee
Otherwise, we obtain
\be
\lambda_m = \lambda_1 \text{.}
\ee
Again, the remaining cases are $\lambda_m = \lambda_3$, which is never better than solution (2) and the limiting case $\lambda_1 = \lambda_3$. \\

As the second example for compression, we assume
\be
\bfsigma = \begin{bmatrix}\lambda_1 & 0 & 0 \\ 0 & -\lambda_2 & 0 \\ 0 & 0 & -\lambda_3\end{bmatrix}\text{,}
\ee
with $\lambda_1, \lambda_2, \lambda_3 > 0$ and we require $\lambda_3 \ge \lambda_1$. The Lagrangian is
\bea
L(\lambda_m,\alpha,\beta) = (\lambda_1-\lambda_m)^2+(-\lambda_2-\lambda_m)^2+ \nonumber \\ (\lambda_m-\lambda_3)^2 - \alpha g(s) - \beta h(t) \rightarrow \underset{\lambda_{m}, \alpha, \beta}{\min} \text{.}
\eea
We calculate
\be
\lambda_{m} = \frac{1}{3} (\lambda_1 - \lambda_2 + \lambda_3) \text{,} \label{eqc1}
\ee
if both of the following inequalities hold 
\be
\lambda_3 - \lambda_2 - 2 \lambda_1 \ge 0 \quad \text{and} \quad 2 \lambda_3 + \lambda_2 - \lambda_1 \ge 0 \text{.} \label{eqc2}
\ee
Otherwise, we calculate 
\be
\lambda_{m} = \lambda_1 \text{.} \label{eqc3}
\ee
The remaining cases result in $\lambda_m \le \lambda_3$, which is never better than solution (2) and the limiting case $\lambda_1 = \lambda_3$.

Lastly, we assume
\be
\bfsigma = \begin{bmatrix}-\lambda_1 & 0 & 0 \\ 0 & -\lambda_2 & 0 \\ 0 & 0 & -\lambda_3\end{bmatrix}\text{,}
\ee
with $\lambda_1 \ge \lambda_2 \ge \lambda_3 > 0$. The Lagrangian is
\bea
L(\lambda_m,\alpha) = (-\lambda_1+\lambda_m)^2+(-\lambda_2-\lambda_m)^2+ \nonumber \\ (-\lambda_3-\lambda_m)^2 - \alpha g(s) \rightarrow \underset{\lambda_{m}, \alpha}{\min} \text{.}
\eea
We calculate
\be
\lambda_m = \frac{1}{3} (\lambda_1 - \lambda_2 - \lambda_3) \text{,}
\ee
if $\lambda_m \le \lambda_1$ holds. Elsewise, we obtain
\be
\lambda_m = \lambda_1 \text{.}
\ee
\subsection{Plane stress}\label{sec:Material4}
As a final problem, we consider a thin sheet of material under plane stress. Without loss of generality we assume the total stress of the material $\bfsigma^t$ equal to zero in all z-directions. This implies that the elastic stress of the material in this direction automatically adjusts to counteract the Maxwell stress to preserve the plane stress state. For our optimization problem this opens up additional possibilities, as the remaining Maxwell stress in the xy-plane (indicated by the following subscript) now has the form
\be
\bftau^m_{xy} = 
\begin{bmatrix}\pm \lambda_m & 0 \\ 0 & - \lambda_m \end{bmatrix} \text{,} \label{eqmwsplane}
\ee
where the sign of the first eigenvalue can now be chosen arbitrarily. Again we use a coordinate system with respect to the principal axes. For the elastic stress, we obtain 
\be
\bfsigma_{xy} = \begin{bmatrix}\lambda_1 & 0 \\ 0 & \lambda_2\end{bmatrix} \text{,}
\ee
where we distinguish between the three cases (1) $\lambda_1 \le \lambda_2 \le 0$, (2) $\lambda_1 \le 0$, $\lambda_2 \ge 0$, (3) $\lambda_2 \ge \lambda_1 \ge 0$. For all cases, we evaluate
\be
||\bfsigma^t||^2 \rightarrow \underset{\lambda_{m}}{\min} \text{,}
\ee
with $||\bfsigma^t||^2  = (\sigma_{xx} + \tau^m_{xx})^2 + (\sigma_{yy} + \tau^m_{yy})^2$. We obtain 
\begin{eqnarray}
\lambda_m =& \frac{1}{2} (-\lambda_1 + \lambda_2) & \quad \text{for case (1) and (2),} \\
\lambda_m =& \frac{1}{2} (\lambda_1 + \lambda_2) & \quad \text{for case (3).}
\end{eqnarray}
Here, for case (1) and (2), we use the positive sign of the first eigenvalue of $\bftau^m_{xy}$ and for case (3) we use the negative sign.
\section{Numerical results}\label{sec:Results}
In this section we show calculations and numerical examples for all cases. As our examples are for illustrative purposes only, we refrain from using specific units.

\begin{figure}[h!]
	\centering
	\includegraphics[width=0.45\textwidth]{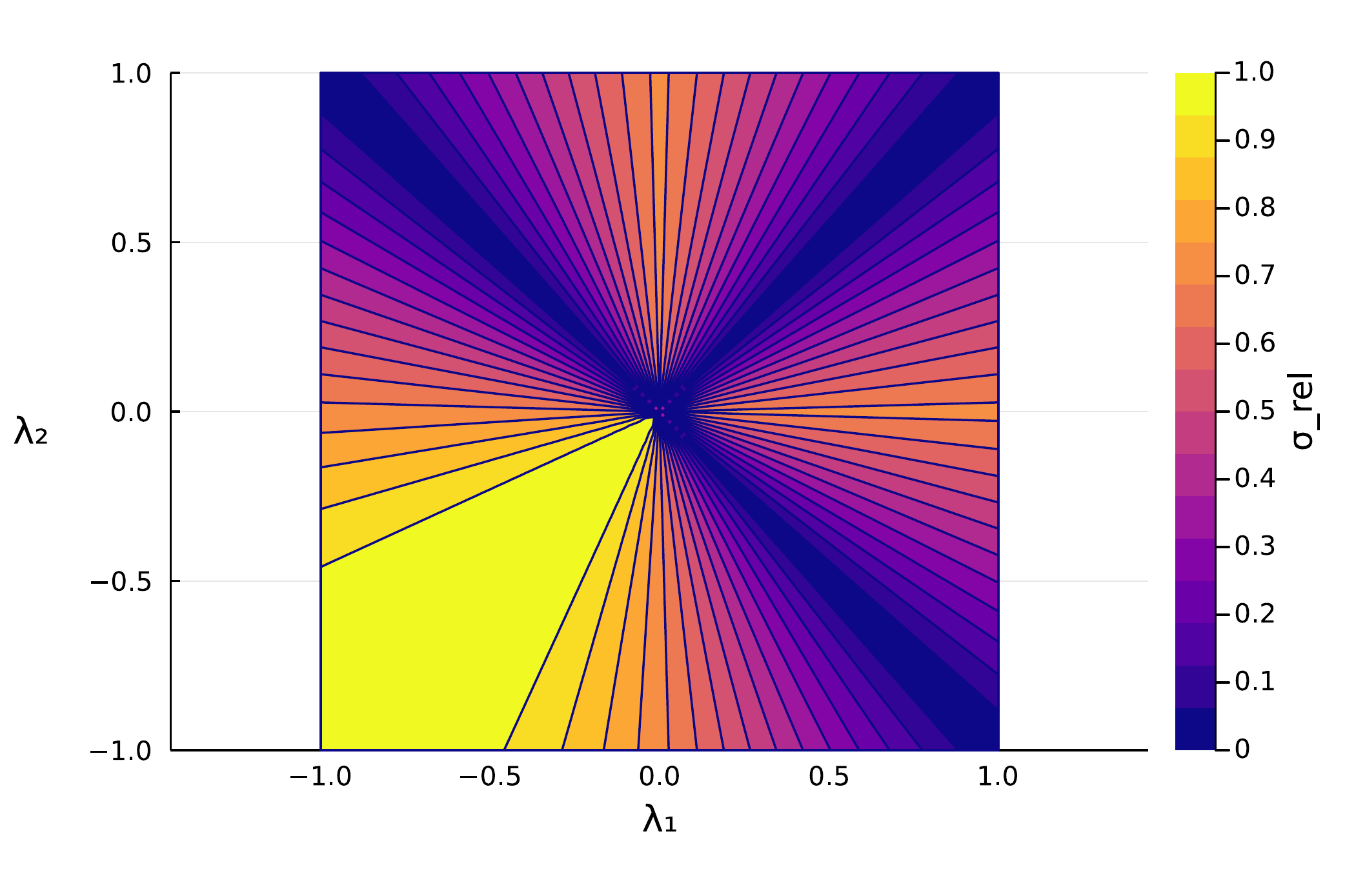}
	\caption{\label{fig1}Remaining stress $\sigma_{\mathrm{rel}}$ depending on the eigenvalues for the plane stress case.}
\end{figure}

We start with the plane stress case, which is simpler to understand and visualize due to its two-dimensional nature. Figure \ref{fig1} shows a map of the remaining stress $\sigma_{\mathrm{rel}}$ depending on the eigenvalues of the stress tensor. 

\begin{figure}[h!]
	\centering
	\includegraphics[width=0.45\textwidth]{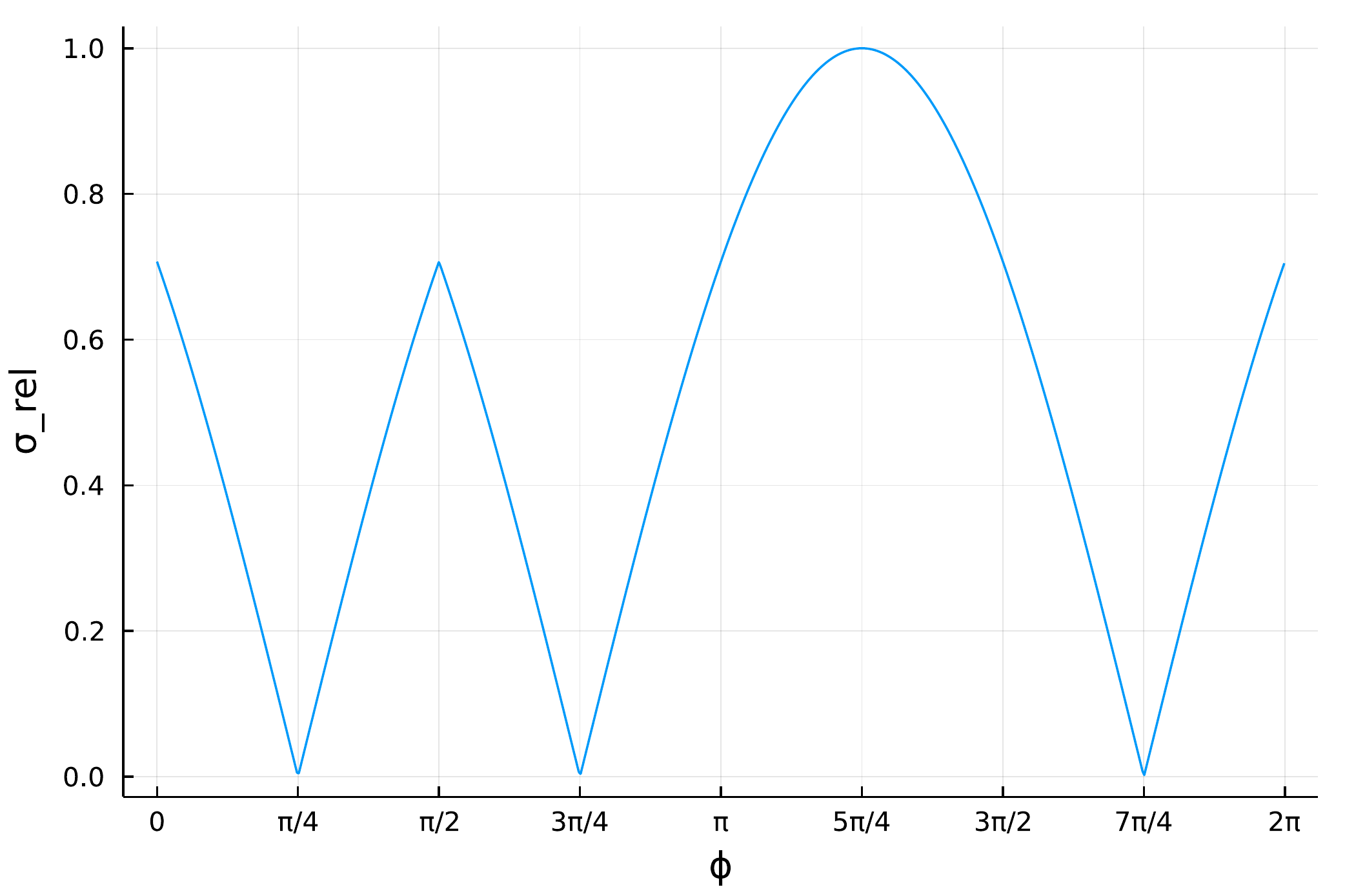}
	\caption{\label{fig2}Remaining stress $\sigma_{\mathrm{rel}}$ depending on the phase angle $\phi$ for the plane stress case.}
\end{figure}

We observe, that from our definition for the relative stress reduction $\sigma_{\mathrm{rel}} = ||\bfsigma + \bftau^m|| / ||\bfsigma||$ (Eq. \eqref{eqsigmarel}), it only depends on the angle with respect to the polar axis (defined as usual as starting at $O = (0, 0)$ and moving horizontally to the right), but not its distance to the origin. Therefore, scaling both eigenvalues with the same factor does not change the result. We investigate this effect further by introducing polar coordinates $(r, \phi)$ with $\lambda_1 = r \cos \phi$, $\lambda_2 = r \sin \phi$, from which follows $\phi = \arctantwo(\lambda_2,\lambda_1)$ and $r = \sqrt{\lambda_{1}^2 + \lambda_{2}^2}$. Using this definition, we calculate the remaining stress as  
\bea
\sigma_{\mathrm{rel}}(\phi) = \begin{cases}
	\frac{\sqrt{2}}{2} \sqrt{1-\sin(2\phi)} & \text{for $\phi \in [0, \frac{\pi}{2}]$}\\
	\frac{\sqrt{2}}{2} |\sin \phi + \cos \phi| & \text{for $\phi \in [\frac{\pi}{2}, 2\pi]$}
\end{cases} \text{,}
 \label{eqsigrelpl}
\eea
which is independent of $r$ as expected. Figure \ref{fig2} shows the resulting plot of this function, which is in conformity with Figure \ref{fig1}. From Eq. \eqref{eqsigrelpl}, we are also able to compute the average remaining stress, which is defined as the mean integral
\be
\bar{\sigma}_{\mathrm{rel}} = \displaystyle\frac{1}{2\pi} \biggl(\displaystyle\int\limits_{0}^{2 \pi} \sigma_{\mathrm{rel}}(\phi) \dd \phi \biggr) \text{.}
\ee
We obtain a solution analytically as 
\bea
\bar{\sigma}_{\mathrm{rel}} &=& \displaystyle\frac{1}{2\pi} \biggl(\displaystyle\int\limits_{0}^{\frac{\pi}{2}} \frac{\sqrt{2}}{2} \sqrt{1-\sin(2\phi)} \dd \phi \nonumber \\
& &+ \displaystyle\int\limits_{\frac{\pi}{2}}^{2 \pi} \frac{\sqrt{2}}{2} |\sin \phi + \cos \phi| \dd \phi \biggr) \nonumber \\
&=& \displaystyle\frac{1}{2\pi}(2-\sqrt{2} + 4 - \sqrt{2}) \approx 0.505 \text{.}
\eea
As can be seen later, compared to the three-dimensional cases, this result is quite good, which can be explained by the signs of the eigenvalues in Eq. \eqref{eqmwsplane}. Case (1) of the plane stress minimization is the only case where it is not possible to choose the signs in such a way that the absolute values of both $\sigma_{xx}$ and $\sigma_{yy}$ are reduced, instead, one value increases the same amount the other one is reduced. This can also be seen in Figure \ref{fig1}, where the bottom-left quadrant yields the worst results compared to the other quadrants. For the other cases, it is very advantageous to choose one sign of the eigenvalues in the Maxwell stress tensor arbitrarily. However, there is a high variance in these results, as there are some rare cases, where the total stress completely vanishes, while in other cases, nearly no stress minimization is possible. It should be noted, that a uniform random creation of all eigenvalues would be biased both for two and three dimensions, therefore, a correct approach, e.g. as proposed in \cite{tashiro1977methods}, should be used. A method for real applications with already known mean stresses was proposed in \cite{gao201718}.

\begin{figure*}[t!]
	\centering
	\includegraphics[width=0.35\textwidth]{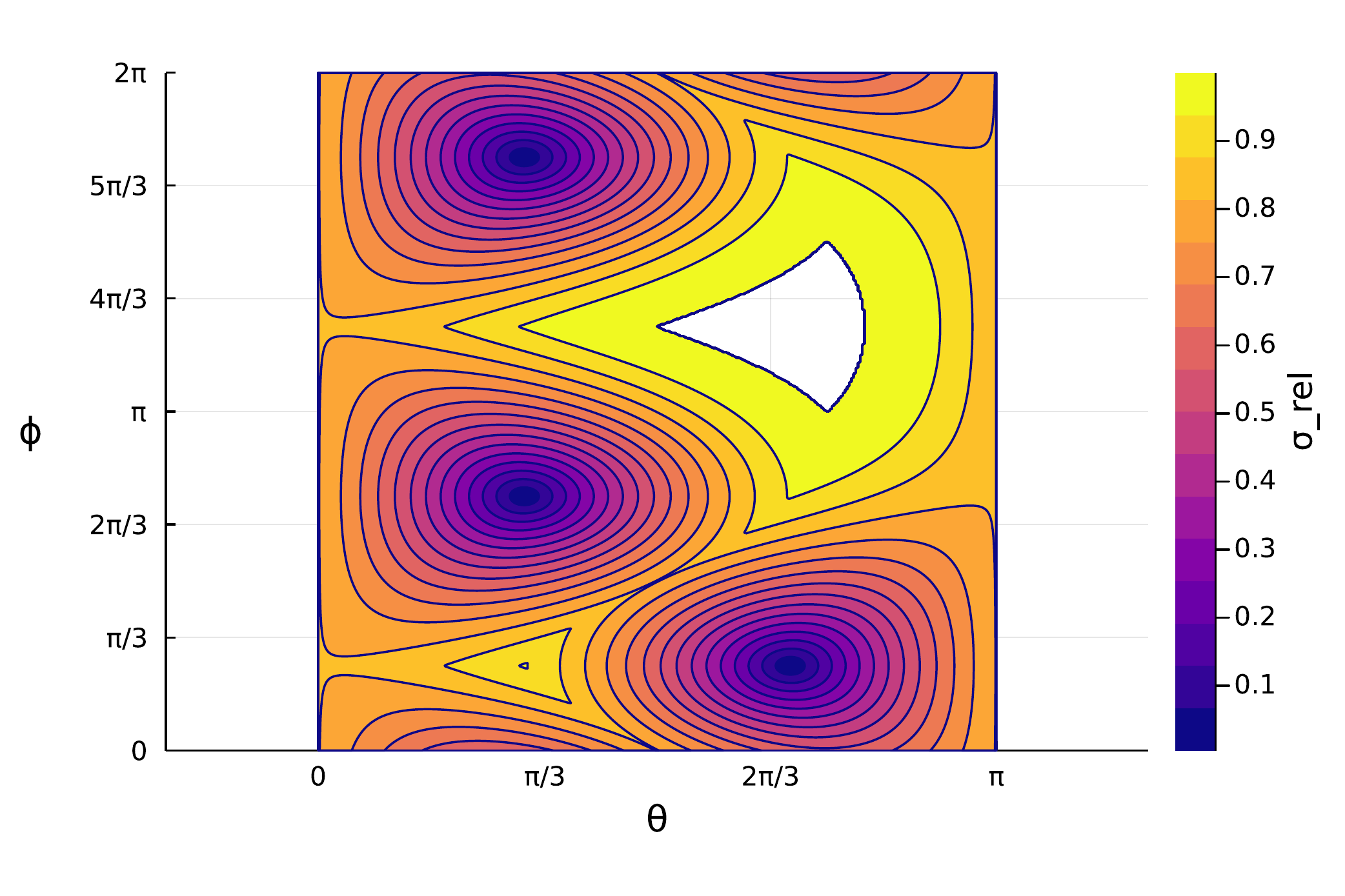} \; \includegraphics[width=0.3\textwidth]{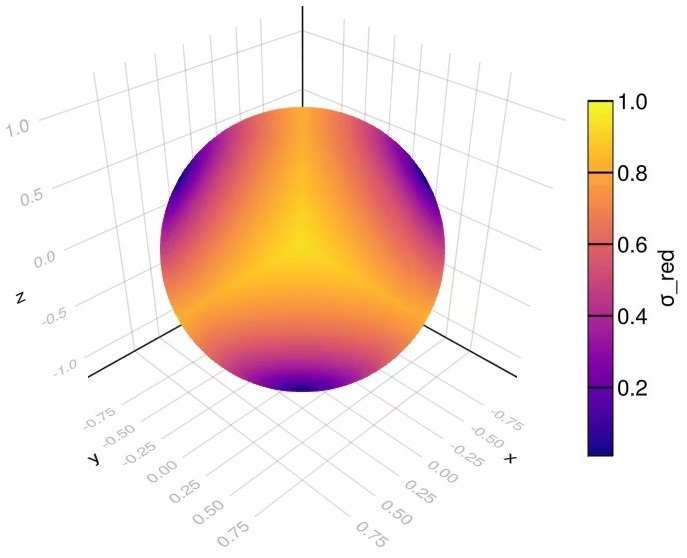} \; \includegraphics[width=0.3\textwidth]{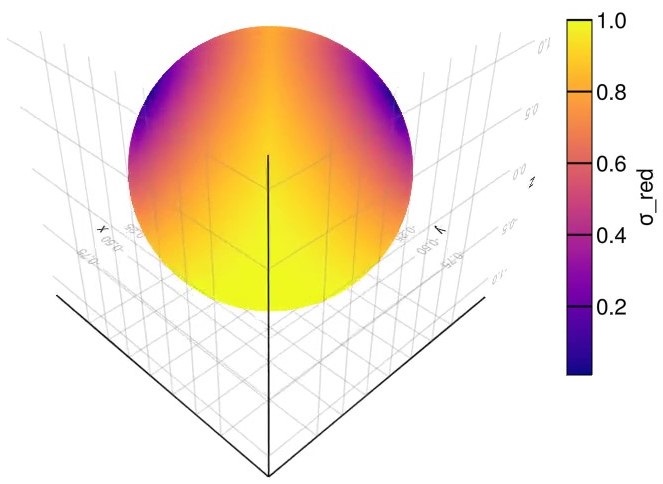}\\
	\includegraphics[width=0.35\textwidth]{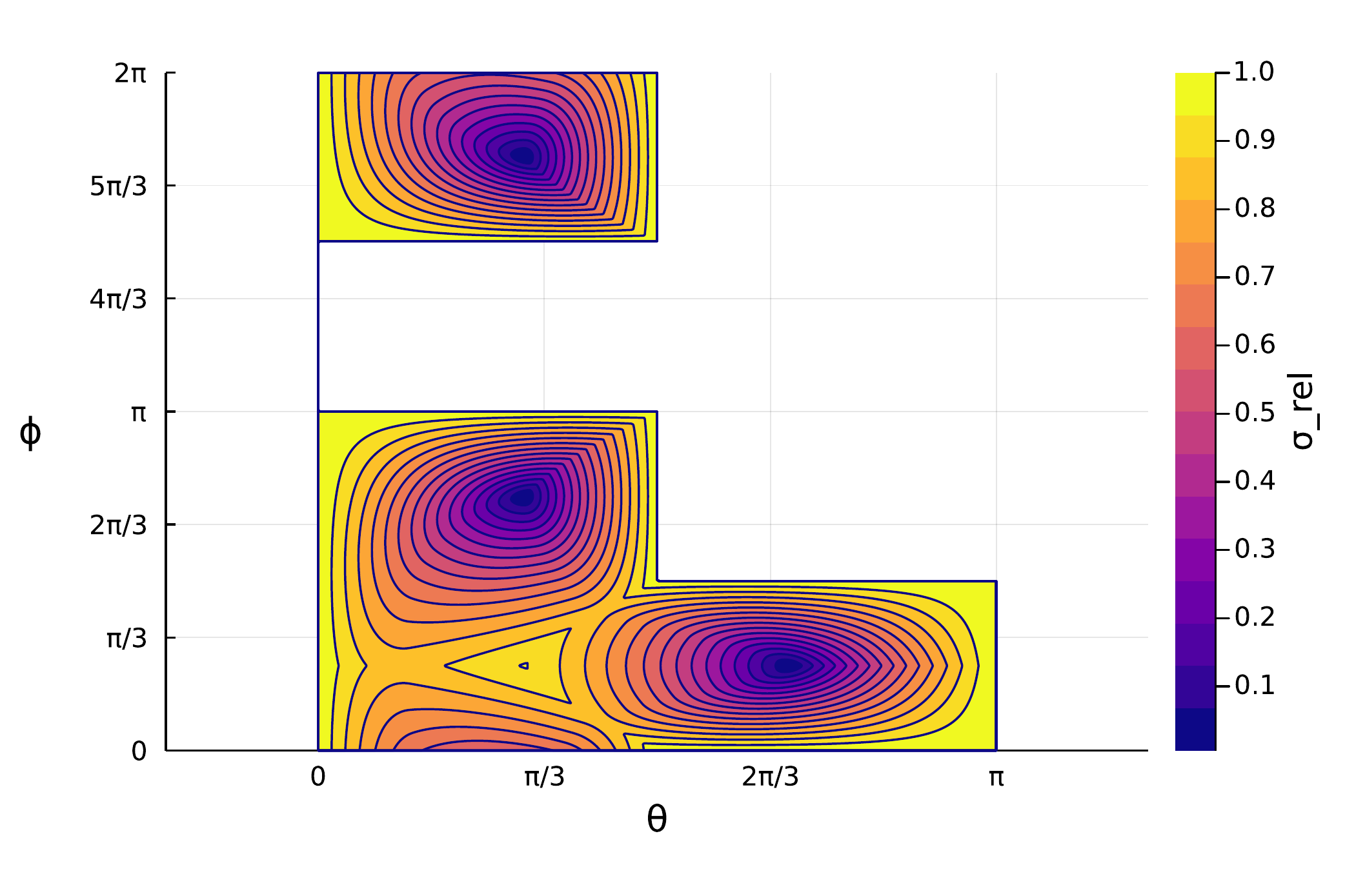} \; \includegraphics[width=0.3\textwidth]{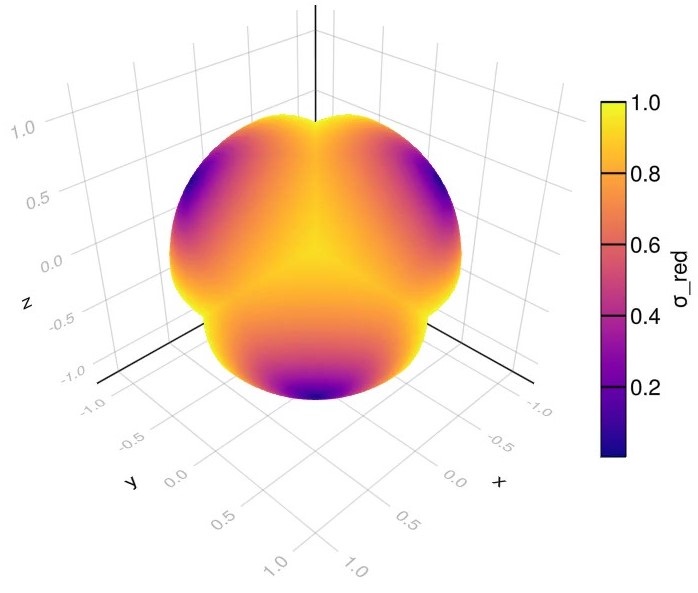} \; \includegraphics[width=0.3\textwidth]{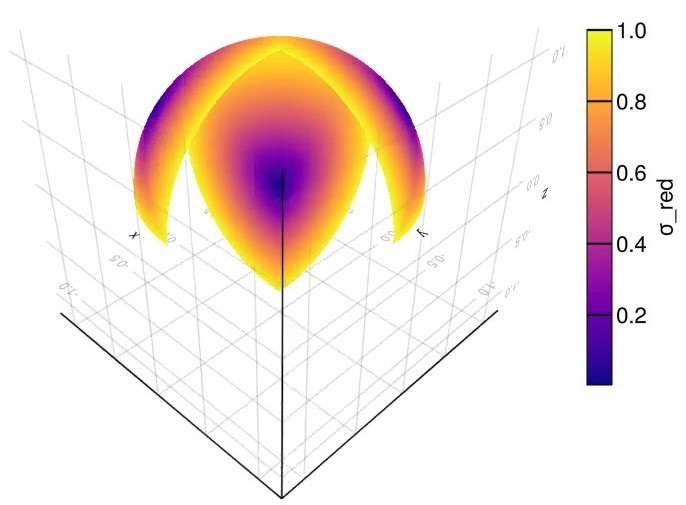}\\
	\includegraphics[width=0.35\textwidth]{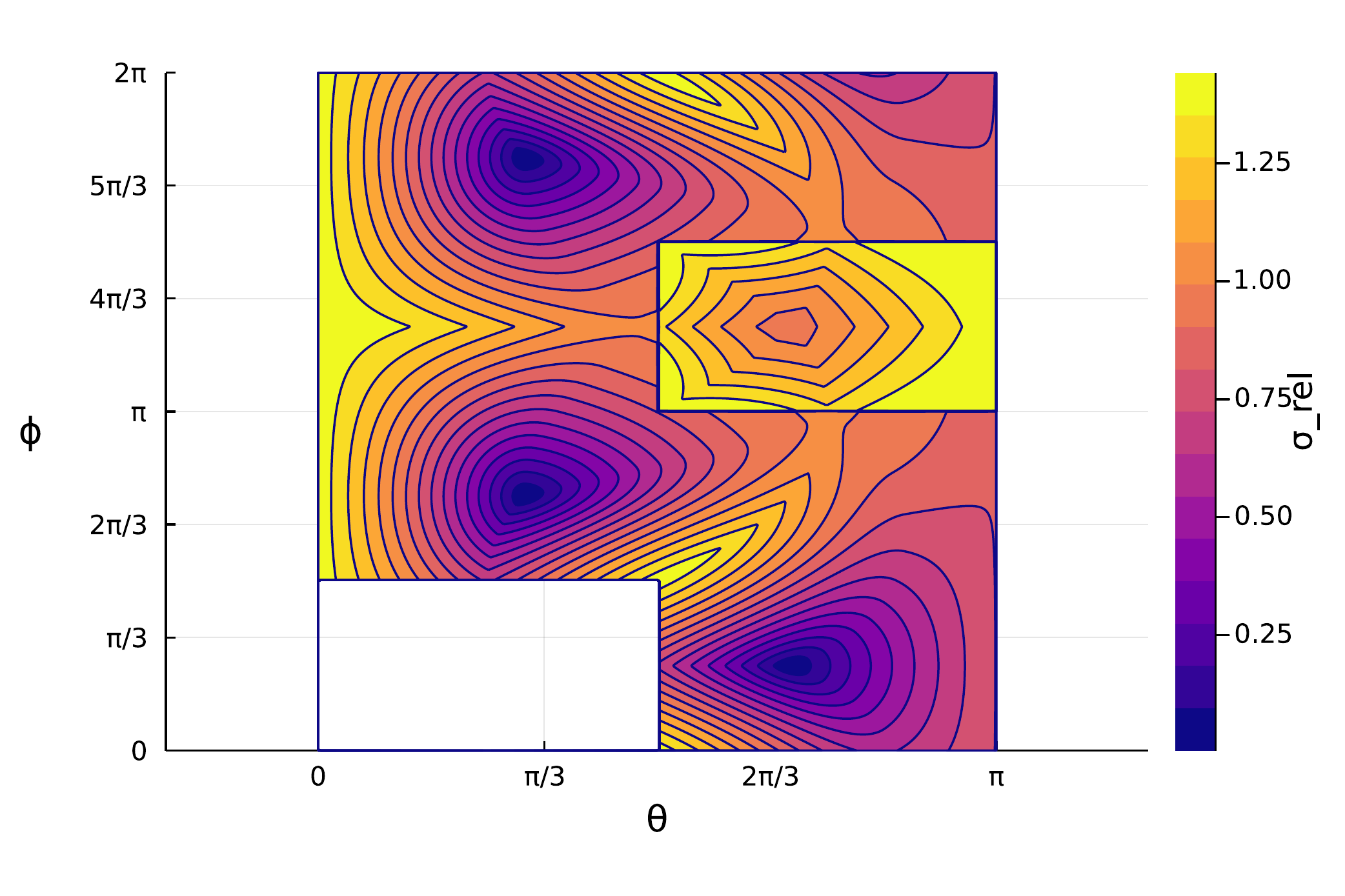} \; \includegraphics[width=0.3\textwidth]{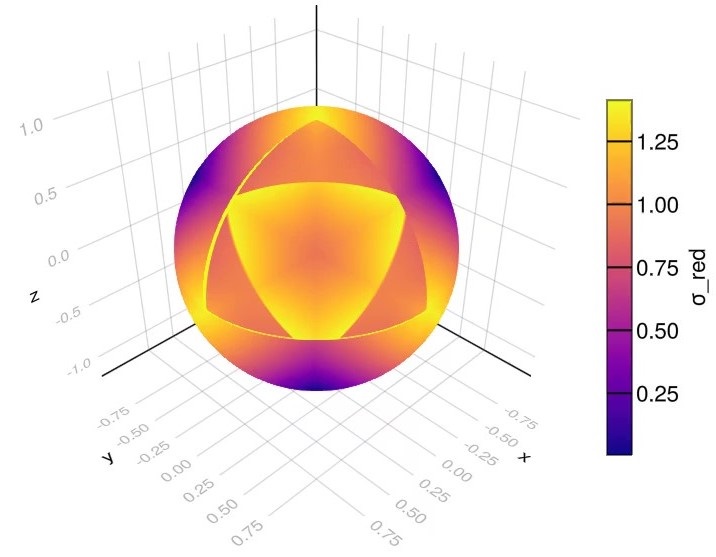} \; \includegraphics[width=0.3\textwidth]{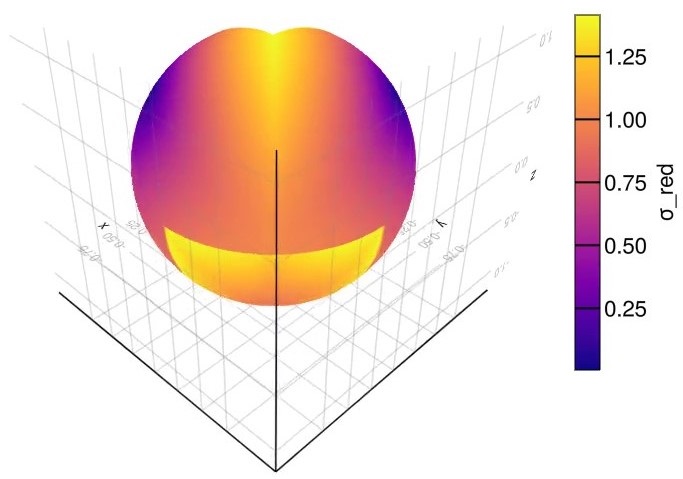}
	\caption{\label{fig3}Left: remaining stress $\sigma_{\mathrm{rel}}$ depending on the polar angle $\theta$ and the azimuthal angle $\phi$ for the three-dimensional cases. in the middle/right: three-dimensional projection depending on the eigenvalues ($x \widehat{=} \lambda_1$, $y \widehat{=} \lambda_2$, $z \widehat{=} \lambda_3$), front side and back side, respectively. Top: unconstrained minimization, in the middle: minimization of tension, bottom: minimization of compression.}
\end{figure*}

As the next step, we evaluate the three-dimensional problems. Analogously to the plane problem, we use the spherical coordinates $(r, \theta, \phi)$, with $\lambda_1 = r \sin\theta \cos\phi$, $\lambda_2 = r \sin\theta \sin\phi$ and $\lambda_3 = r \cos\theta$, from which follows $\phi  = \arctantwo(\lambda_2, \lambda_1)$ and $\theta = \arccos(\frac{\lambda_{3}}{r})$, with $r = \sqrt{\lambda_{1}^2 + \lambda_{2}^2 + \lambda_{3}^2}$. By inserting this into our calculation of the relative stress $\sigma_{\mathrm{rel}}$ (Eq. \eqref{eqsigmarel}) we can again show that it is independent of the radius $r$. Figure \ref{fig3} shows the results for the three-dimensional minimization problems.
For the unconstrained minimization (top row), we obtain three areas, where nearly a complete stress reduction is possible. Here, exactly two of the three eigenvalues are positive and all eigenvalues have the same or similar absolute values. On the other hand, there is a significant void area in the section for three negative eigenvalues, where no solution can be found, as seen in Eq. \eqref{eqalpha}. In contrast to the previous example, the resulting functions for the different cases do not have an analytical primitive function. Therefore, we solve the problem numerically by randomly creating a sample of $100000$ uniformly distributed points on the unit sphere. The angles for the creation have the distributions  $\phi \sim 2 \pi \; \mathcal{U}[0,1]$ and $\theta \sim \arccos{(1-2 \; \mathcal{U}[0,1])}$ \cite{weisstein2002sphere}. Then, averaging the result of our stress reduction gives $\bar{\sigma}_{\mathrm{rel}} = 0.70$, where we assumed ${\sigma}_{\mathrm{rel}} = 1$ for the cases without a real solution. We repeat the procedure for the tension minimization problem, whose results depending on the two angles are plotted in the middle row of Figure \ref{fig3}.

Again, we obtain the three areas with a very low remaining stress at the same positions, but the rate, at which $\sigma_{\mathrm{rel}}$ increases moving away from these points is higher. This time, the size of the void area is half of the total size and contains all cases with at least two negative eigenvalues, for which a minimization under the tensile constraint is not possible. For the calculation of the average remaining stress, we exclude these areas and calculate $\bar{\sigma}_{\mathrm{rel}} = 0.66$ by again using our uniform distribution. 

Finally, the bottom row of Figure \ref{fig3} shows the results for the compression problem. Here, only for the part where all eigenvalues are positive, no solution can be found. Similarly to the other examples we obtain the three areas, where a very good stress absorption is possible. For the compression problem, it is very noticeably that $\sigma_{\mathrm{rel}}$ abruptly increases in the area with three negative eigenvalues. In order to fulfill the compression constraint, there are cases where the resulting stress is higher than before ($\sigma_{\mathrm{rel}} > 1$). From the randomization, we calculate the average stress reduction $\bar{\sigma}_{\mathrm{rel}} = 0.87$, again excluding the area, where no solution can be found.
\section{Discussion}\label{sec:Discussion}

While our calculations seem promising, some limitations occur. To put our approach into practice, the unit cells have to be small enough for the mechanical stress to be assumed constant within the cell. Here, the absolute size may differ depending on the application. We focused on microscale calculations, but the interference of the electric field from neighboring unit cells could be an important effect to consider. As shown by our calculations, a complete absorption of the mechanical stress is only possible for rare special cases, however a significant stress reduction is possible for most cases. The variance of the stress reduction is very high and the result depends on the specific stress state. Furthermore, the maximum absolute values of reducible stresses depend on the technical possibilities of the used capacitors and the used bulk material, as a very high voltage may cause an dielectric breakdown. Therefore, possible applications could be found at smaller scales, e.g. robotic capsules for endoscopy or drug delivery in medical engineering \cite{cortegoso2021overview}. To automate the process and allow the usage for not only static but also dynamic problems or applications, where the material load is not known beforehand, sensor technology plays an important role. In order to counteract the mechanical stress, the electric field has to be calculated and applied quickly from sensor measurements, especially if the load changes fast in time. For further development, we suggest experimental research of our proposed material. Finally, a parameter study evaluating the influence of the relative permittivity $\varepsilon_r$ could be done.
\bibliographystyle{unsrt}
\bibliography{references}
\end{document}